\documentclass[aps,showpacs,superscriptaddress,groupedaddress]{revtex4}
\usepackage{graphicx}
\usepackage{amsmath}
\usepackage{amssymb}
\usepackage{slashed}
\usepackage[compat=1.0.0]{tikz-feynman}

\tikzfeynmanset{
	fermion2/.style={
		/tikz/postaction={
			/tikz/decorate=false,
		},
	},
}
\tikzfeynmanset{
	fermion3/.style={
		/tikz/decoration={
			markings,
			mark=at position 0.5 with {
				\node[
				dot,
				fill,
				draw=none,
				] { };
			},
		},
		/tikz/postaction={
			/tikz/decorate=true,
		},
	},
}

\DeclareMathOperator*{\SumInt}{%
	\mathchoice%
	{\ooalign{\raisebox{.15\height}{\scalebox{0.9}{$\textstyle\sum$}}\cr\hidewidth$\displaystyle\int$\hidewidth\cr}}
	{\ooalign{\raisebox{.14\height}{\scalebox{.7}{$\textstyle\sum$}}\cr\hidewidth$\textstyle\int$\hidewidth\cr}}
	{\ooalign{\raisebox{.2\height}{\scalebox{.6}{$\scriptstyle\sum$}}\cr$\scriptstyle\int$\cr}}
	{\ooalign{\raisebox{.2\height}{\scalebox{.6}{$\scriptstyle\sum$}}\cr$\scriptstyle\int$\cr}}
}

\newcommand{\avg}[1]{\left\langle #1 \right\rangle}

\newcommand{\Tr}{\textrm{Tr}}

\newcommand{\Eq}[1]{Eq.~(\ref{#1})}
\newcommand{\Eqs}[1]{Eqs.~(\ref{#1})}
\newcommand{\vx}{\vec{x}}
\newcommand{\vq}{\vec{q}}
\newcommand{\vk}{\vec{k}}
\newcommand{\vp}{\vec{p}}
\newcommand{\vzero}{\vec{0}}

\definecolor{dgreen}{rgb}{0.1,0.5,0.1}
\definecolor{lblue}{rgb}{0.2,0.35,1}
\definecolor{webred}{rgb}{0.75,0,0}

\begin{document}
	
	\title{Towards a Stability Analysis of Inhomogeneous Phases in QCD}
	
	\author{Theo F. Motta}
	\affiliation{Institut für Theoretische Physik, Justus-Liebig-Universität Gie\ss en, 35392 Gie\ss en, Germany.\\ \,}
	\affiliation{Technische Universität Darmstadt, Fachbereich Physik, Institut für Kernphysik, Theoriezentrum, Schlossgartenstr.\ 2 D-64289 Darmstadt, Germany.\\ \,}
	
	\author{Julian Bernhardt}
	\affiliation{Institut für Theoretische Physik, Justus-Liebig-Universität Gie\ss en, 35392 Gie\ss en, Germany.\\ \,}

	\author{Michael Buballa}
	\affiliation{Technische Universität Darmstadt, Fachbereich Physik, Institut für Kernphysik, Theoriezentrum, Schlossgartenstr.\ 2 D-64289 Darmstadt, Germany.\\ \,}
	\affiliation{Helmholtz Forschungsakademie Hessen f\"{u}r FAIR (HFHF), \\
		GSI Helmholtzzentrum f\"{u}r Schwerionenforschung,\\
		Campus Darmstadt,
		64289 Darmstadt,
		Germany. \\ \,}
	
	\author{Christian S. Fischer}
	\affiliation{Institut für Theoretische Physik, Justus-Liebig-Universität Gie\ss en, 35392 Gie\ss en, Germany.\\ \,}
	\affiliation{Helmholtz Forschungsakademie Hessen f\"{u}r FAIR (HFHF), \\
		GSI Helmholtzzentrum f\"{u}r Schwerionenforschung,\\
		Campus Gie\ss{}en,
		35392 Gie\ss{}en,
		Germany.}

	\begin{abstract}
		The possible occurrence of crystalline or inhomogeneous phases in the QCD phase diagram at 
		large chemical potential has been under investigation for over thirty years. Such phases are 
		present in \textit{models} of QCD such as the Gross--Neveu model in 1+1 dimensions, 
		Nambu--Jona-Lasinio  (NJL) and quark meson models. Yet, no unambiguous confirmation exists 
		from actual QCD. In this work, we propose a new approach for a stability analysis that is 
		based on the two-particle irreducible effective action and compatible with full QCD calculations 
		within the framework of functional methods. As a first test, we reproduce a known NJL model 
		result within this framework. We then discuss the additional difficulties which arise in QCD 
		due to the non-locality of the quark self-energy and suggest a method to overcome them. 
		As a proof of principle and as an illustration of the analysis, we consider the Wigner-Weyl 
		solution of the quark Dyson-Schwinger equation (DSE) within a simple truncation of QCD 
		in the chiral limit and analyse its stability against {\it homogeneous} chiral-symmetry 
		breaking fluctuations. For temperatures above and below the tricritical point we find that 
		the boundary of the instability region coincides well with the second-order phase boundary 
		or the left spinodal, respectively, obtained from the direct solutions of the DSEs. 
		Finally, we outline how this method can be generalized to study inhomogeneous fluctuations.
		
	\end{abstract}
	
	\maketitle
	
	\section{Introduction}
	At high densities and low temperatures, systems of interacting fermions may form crystalline, i.e., spatially inhomogeneous phases. Unsurprisingly,  systems governed by the strong interaction are no exception. 
	The idea of density waves in nuclear matter dates back to Overhauser in 1960 \cite{Overhauser:1960},
	followed by Migdal's renowned works on p-wave pion condensation in the 1970s \cite{Migdal:1973zm,Migdal:1978az}.
	A first relativistic treatment of chiral-density waves (CDWs) in nuclear matter was presented in Ref.~\cite{Dautry:1979bk},
	while already in 1990 the idea was transferred to quark matter \cite{Kutschera:1989yz} (see Ref.~\cite{Broniowski:2011ef} for a brief review on those early works). 
	In the early 2000s, the possible existence of colour superconducting (CSC) phases attracted much attention \cite{Alford:2007xm},
	and in this context both crystalline CSC phases \cite{Alford:2000ze,Bowers:2001ip,Leibovich:2001xr,Giannakis:2002jh,Mannarelli:2006fy,Cao:2015rea,Anglani:2007aa}
	as well as inhomogeneous chiral phases in coexistence with homogeneous CSC phases \cite{Sadzikowski:2002iy,Sadzikowski:2006jq} 
	have been explored. Roughly at the same time, sophisticated analyses of the 1+1 dimensional Gross Neveu (GN) and chiral GN models in the large-$N$ limit revealed not only the existence of inhomogeneous phases in these models but also the exact shapes 
	of the inhomogeneities \cite{Schon:2000qy,Thies:2006ti,Basar:2009fg,Ciccone_2022}. The discovery that these lower-dimensional solutions could also be embedded into higher dimensions \cite{Nickel:2009wj} triggered further studies in particular in the Nambu--Jona-Lasinio (NJL) and Quark-Meson (QM) models, which seemed to confirm that inhomogeneous phases 
	are a rather robust feature of this kind of models (see Ref.~\cite{Buballa:2014tba} for a review). 
	
	However, while the existence of inhomogeneous phases has rigorously been proven for the $1+1$ dimensional GN and chiral GN models in the large-$N$ limit, recent works within higher dimensional NJL and GN models signal the possibility that inhomogeneous phases may only appear as cutoff artefacts  \cite{Buballa:2020nsi,Pannullo:2022eqh,Pannullo:2023one}, also see Refs. \cite{Broniowski:1990gb,Partyka:2008sv} for early studies on cutoff effects. This casts  some general doubt that models that are non-renormalisable and, consequentially, cutoff dependent, could ever produce trustworthy results concerning such phases. 
	
	It is thus an interesting and open question, whether inhomogeneous phases exist in actual Quantum Chromodynamics (QCD). 
	In fact, it was shown in Refs.~\cite{Deryagin:1992rw,Shuster:1999tn} that, for large number of quark colours $N_c$, homogeneous quark matter at high density is unstable against the formation of chiral waves -- a phase in which chiral symmetry is inhomogeneously broken\footnote{Inhomogeneous chiral waves at large $N_c$ have also been discussed in the context of quarkyonic matter \cite{Kojo:2009ha}.}. Yet Ref.~\cite{Shuster:1999tn} also showed that this was unlikely to happen at $N_c=3$, for which they are strongly disfavored against CSC phases.
	These analyses, however, rely on weak-coupling expansions, which are valid only at very high densities.
	They are therefore not applicable to the phenomenologically more interesting nonperturbative regime between about one  
	and a few times nuclear saturation density.
	
	On the other hand, since this regime is not accessible by lattice simulations either, the question about the existence of inhomogeneous
	chiral phases in QCD has to be addressed with functional methods. 
	In Ref.~\cite{Fu:2019hdw}, where  three-flavour QCD was investigated within the functional renormalisation group,
	indications of inhomogeneity in a region around the critical endpoint have been observed but the finding was not fully conclusive.\footnote{The authors identified a so-called moat regime where the wave-function renormalisation constant in the scalar channel is negative.
		This is a necessary but not sufficient condition for instability of the homogeneous phase against small inhomogeneous fluctuations.}
	Within a truncation of the (renormalised) Dyson-Schwinger Equations (DSE) of QCD Ref.~\cite{Muller:2013tya} managed to find an inhomogeneous solution to the quark propagator at finite density. 
	However, the analysis was restricted to a particular ansatz for the inhomogeneity, making it extremely difficult to improve the truncation. 
	
	In this work, we therefore aim at developing a more flexible approach to study the existence of inhomogeneous phases in QCD.
	We propose a generalization of the stability analyses that have successfully been applied to models like NJL
	\cite{Pannullo:2022eqh,Buballa:2018hux,Carignano:2019ivp}, GN \cite{Buballa:2020nsi,Koenigstein:2021llr}
	or the QM \cite{Buballa:2020xaa} model.
	In its new form the method is not only applicable to renormalisable theories such as QCD but also to beyond mean-field applications of NJL and QM models.
	
	The paper is organised as follows: 
	In order to place our work in the context of previous approaches, we briefly review in section \ref{sec:models}
	how inhomogeneous phases are studied within models, putting particular emphasis on stability analyses in the NJL model.
	As we will discuss, this method cannot immediately be applied to to QCD. 
	In section \ref{sec:staba} we will therefore present a new framework using the 2PI effective action of QCD. We then follow with
	a small number of general remarks on potential generalisations to nPI effective actions. 
	In section \ref{sec:NJL}, as a first test, we apply the approach again to the NJL model and show that we can reproduce the known results.
	Then, in section \ref{sec:fullsa} we discuss additional difficulties which arise in QCD as compared to the NJL model and propose a
	method how to overcome them. 
	In section \ref{sec:chiral} we perform a first numerical test of  this approach by applying it to study the second-order transition from 
	the {\it homogeneous} chirally broken to the chirally restored phase in the chiral limit of (a truncation of) QCD 
	and comparing the result to that from directly solving the corresponding DSE.
	As we will see, this gives vital clues as to the working of the method and how to extend it to study inhomogeneous phases.
	We conclude this work with a short summary and an outlook towards more sophisticated calculations within QCD.

	\section{Model Studies of Inhomogeneous Phases}
	\label{sec:models}
	As mentioned above, most studies of inhomogeneous phases in strong-interaction matter have been performed within 
	QCD inspired models. 
	In addition simplifications have been used, since it is a non-trivial matter to study phases which break translational symmetry. 
	Most commonly, two approaches are available. First, one might propose an \textit{ad hoc} ansatz for an inhomogeneous phase and simply 
	verifies whether or not this proposed ansatz phase is more stable than the homogeneous one (by comparing their free energies).
	Alternatively, one may perform a stability analysis of the homogeneous phase, i.e. one determines whether 
	or not the homogeneous phase is unstable against {small} inhomogeneous fluctuations of \textit{any} shape. Both NJL, GN and QM 
	studies have been carried out within these two approaches (see, e.g.
	\cite{Carignano:2019ivp,Koenigstein:2021llr,Buballa:2020xaa,Lakaschus:2020caq,Carignano:2018hvn,Carignano:2017meb} and references therein). 
	
	As an illustration, take the standard NJL model
	\begin{equation}
	\label{LNJL}
	\mathcal{L}_\text{NJL}=\bar{\psi}(i \slashed \partial-m) \psi
	+G \sum_M \big(\bar{\psi} V^{(M)} \psi\big)^2\,
	\end{equation}
	where $\psi$ denotes a quark field with two flavour degrees of freedom and bare mass $m$.
	The second term with coupling constant $G>0$ and vertices  $V^{(M)} \in \{\boldsymbol{1}, i\gamma_5 \tau_a\}$, where  
	$\tau_a$, $a=1,2,3$, are the Pauli matrices in isospin space, correspond to four-point interactions in the scalar-isoscalar and 
	pseudoscalar-isovector channels.
	Assuming the presence of the (possibly inhomogeneous) condensates
	\begin{equation}
	\phi_M(\vx)=\big\langle\bar{\psi}(\vx) V^{(M)} \psi(\vx)\big\rangle,		
	\end{equation}
	the mean-field thermodynamic potential, corresponding to the free-energy density, then takes the form
	\begin{equation}\label{freeE}
	\begin{aligned}
	\Omega_{\mathrm{MF}}=&-\frac{T}{V} \operatorname{Tr} \log \left(\frac{S^{-1}}{T}\right)+G \sum_M \frac{1}{V} \int d^3 x\, \phi_M^2(\vx) ,
	\end{aligned}
	\end{equation}
	where
	\begin{equation}
	\label{Sinv}
	S^{-1} = S_0^{-1} 
	+ 2 G \sum_M V^{(M)}  \phi_M(\vx)\
	\end{equation}
	is the inverse dressed quark propagator and $S_0^{-1}$ is its bare counterpart. 
	$T$ is the temperature, while the chemical potential is hidden in the bare propagator. 
	The functional trace $ \operatorname{Tr}$ is over internal degrees of freedom as well as the Euclidean four-volume 
	$V_4=[0,\frac{1}{T}] \times V$ with the spatial volume $V$, which will be sent to infinity.
	
	It is important to note that the quark self-energy $\Sigma=S^{-1} - S_0^{-1}$ is a linear combination of the condensates.
	This is a feature of mean-field NJL, which is absent in QCD. It allows us to view the mean-field thermodynamic potential
	as a functional of the condensates, $\Omega_{\mathrm{MF}} = \Omega_{\mathrm{MF}}[\phi_M]$. 
	Yet, in order to determine the ground state of the system, we have to minimize $\Omega_{\mathrm{MF}}$ with respect to the 
	condensates, which is a nontrivial task as their spatial shapes are unknown.\footnote{Strictly speaking, the thermodynamic potential
		is $V\Omega_{\mathrm{MF}}$ evaluated at this ground-state configuration, while the functional 
		$\Omega_{\mathrm{MF}}[\phi_M]$ corresponds to an effective potential per volume. As customary, however, we call 
		$\Omega_{\mathrm{MF}}[\phi_M]$ thermodynamic potential as well.}

	In this context, basically two strategies have been employed in the literature.
	The first one consists in making a direct ansatz for the condensates with a finite number of parameters and then minimizing the free
	energy with respect to these parameters. Probably the most popular example is the Chiral Density Wave (CDW) \cite{Nakano:2004cd,Dautry:1979bk,Kutschera:1989yz},
	\begin{equation}
	\phi_S(\vx) = -\frac{\Delta}{2G}\cos(\vq\cdot \vx), \quad
	\phi_{P,3}(\vx) = -\frac{\Delta}{2G}\sin(\vq\cdot \vx),
	\end{equation}
	with parameters $\Delta$, related to the amplitude, and a wave vector $\vq$.
	This ansatz corresponds to a rotation in the plane of scalar ($V^{(S)} =  \boldsymbol{1}$)
	and pseudoscalar ($V^{(P,3)} = i\gamma_5 \tau_3$) condensates
	as one goes along the direction of the wave vector $\vq$.
	A more sophisticated ansatz is the Real-Kink-Crystal (RKC) \cite{Nickel:2009wj}, 
	where the scalar condensate takes the form of a Jacobi elliptic function,
	$\phi_S(x) = \Delta \sqrt{\nu}/2G \ \text{sn}(\Delta x | \nu)$, with elliptic modulus $\nu$.   
	Other examples studied are one- and two-dimensional sinusoidal shapes in the scalar sector \cite{Carignano:2012sx}.
	All of these shapes were typically found to be favoured over the homogeneous phase in a certain region of the phase diagram,
	with the RCS ansatz being the most favoured one studied so far \cite{Buballa:2014tba}.
	
	The alternative approach is performing a stability analysis of the homogeneous ground state, i.e., the homogeneous state
	with the lowest free energy. 
	In contrast to the direct ansatz approach it has the advantage of being ``modulation shape agnostic''. 
	This may be a real benefit, since in practice we usually do not know the exact shape the inhomogeneous modulation takes.\footnote{Remarkable exceptions are the $1+1$ dimensional non-chiral and chiral Gross-Neveu models, where the inhomogeneous
		solutions have been constructed explicitly, namely the RKC and the CDW (``chiral spiral''), respectively~\cite{Thies:2006ti}.} 
	On the downside, since
	the analysis relies on continuity, discontinuous first-order transitions to inhomogeneous phases may not show up in a stability analysis.
	
	Specifically, one expands the free energy in Eq.~(\ref{freeE}) around the homogeneous state. 
	The condensates are then expressed as
	\begin{equation}
	\phi_M(\vx)=\bar\phi_{M}+\delta \phi_M(\vx), 
	\end{equation}
	where in this paper we indicate quantities in the homogeneous ground state with a bar. 
	So $\bar\phi_{M}$ denotes the condensates in the homogeneous ground state while the $\delta \phi_M(x)$ are small, possibly inhomogeneous, perturbations around them.
	We then expand
	\begin{equation}
	\Omega_{\mathrm{MF}}=\sum_{n=0}^{\infty} \Omega^{(n)},
	\qquad \Omega^{(n)} \propto \mathcal{O}(\delta \phi^n)
	\end{equation}
	where each contribution $\Omega^{(n)}$ is of the $n$-th power in the fluctuations.
	Obviously, the zeroth order is the homogeneous case
	\begin{equation}
	\Omega^{(0)}_\text{NJL} = -\frac{T}{V} \operatorname{Tr} \log \left(\frac{\bar S^{-1}}{T}\right)+G\sum_M \bar{\phi}_M^2,
	\end{equation}
	where ${\bar S^{-1}}$ is the inverse quark propagator \Eq{Sinv} evaluated for $\phi_M = \bar\phi_M$.
	
	The first-order term $\Omega^{(1)}$ vanishes by the quark gap equation \cite{Buballa:2018hux}, and therefore the leading-order
	fluctuation contribution is the second-order term, which can be written as
	\begin{equation}
	\label{classicNJLcondition}
	\Omega^{(2)}  
	= \frac{2 G^2}{V}  \sum_M \int\limits_{\vq} \, |\delta\phi_M(\vq)|^2 D_M^{-1}(q)
	\end{equation}
	where the $ \delta\phi_M(\vq)$ is the Fourier transform\footnote{Here we work with continuous Fourier transforms (see \ref{appendixFourier})
		for convenience. 
		On the other hand, typical inhomogeneities which have been discussed in this context are periodic in space, 
		corresponding to a discrete reciprocal lattice (R.L.) in Fourier space. 
		This is contained in our formalism as $\delta\phi_M(\vq) = \sum_{{\vq}_n \in R.L.} 
		\delta(\vq - {\vq}_n) \,\delta\phi_{M,{\vq}_n}$
		with discrete Fourier coefficients $\delta\phi_{M,{\vq}_n}$.
		Using $[\delta(\vq)]^2 = \delta(\vq)\delta(\vzero) =  \delta(\vq) \int d^3x = V \delta(\vq)$, 
		the squared Fourier transform in \Eq{classicNJLcondition} should then be replaced by
		$ |\delta\phi_M(\vq)|^2 \rightarrow V  \sum_{{\vq}_n \in R.L.} \delta(\vq - {\vq}_n) \,|\delta\phi_{M,{\vq}_n}|^2$. 
		The volume $V$ then drops out and \Eq{classicNJLcondition} becomes equal to the result presented
		in Ref.~\cite{Buballa:2018hux} and others.}
	of $\delta\phi_M(\vx)$ and
	\begin{equation}
	\label{DM}
	D_M^{-1}(q) =
	\frac{1}{2G} + \SumInt_{k}\text{tr}\big[ V^{(M)} \bar{S}(k) V^{(M)} \bar{S}(k+q) \big]
	\end{equation}
	can be interpreted as inverse (unrenormalized) meson propagators within the NJL model. 
	Here ``tr'' denotes a trace over internal (colour, flavour, and Dirac) degrees of freedom,
	and we introduced the short-hand notations
	\begin{equation}\label{sumint}
	\int\limits_{\vk} \equiv \int \frac{d^3k}{(2\pi)^3} \,, 
	\quad 
	\SumInt_{k} \equiv T\sum_{\omega_n}\int \frac{d^3 k}{(2\pi)^3}
	\end{equation}
	where momenta are generically given by $k = (\vec{k},k_4=\omega_k)$, with fermionic Matsubara frequencies $\omega_k = (2n+1)\pi T$.
	
	Note that, since we only allow for spatial inhomogeneities but still assume that the fluctuations are time-independent,
	$\delta\phi_{M}(\vq)$ only depends on the three-momentum $\vq$, which is integrated over in  \Eq{classicNJLcondition}.  
	The argument $q$ of $D_M^{-1}$ is then the corresponding four-momentum vector with zero energy. 
	Moreover, since we are expanding about a homogeneous and therefore isotropic state, $D_M^{-1}$ effectively only depends on $|\vq|$.
	
	An important feature of \Eq{classicNJLcondition} is that the integrand depends on the squared modulus of $\delta\phi_{M}(\vq)$, so the only way this can ever be negative, and thus, lower the free energy, is by having the inverse meson propagators $D_M^{-1}(q)$ be negative for certain momenta. This is actually a sufficient criterion for instability because, even if $D_M^{-1}(q)$ is negative only in a small momentum
	regime, any fluctuation which has support only in that regime will lower the free energy. 
	The phase boundary can then be identified as a line in the $T$-$\mu$ plane where one of the $D_M^{-1}$ touches zero at a single value
	of $|\vq|$.
	
	As obvious from \Eq{DM}, in the limit of vanishing interactions the positive constant $1/2G$ dominates and no instability occurs.
	On the other hand, the integral is negative, as we will discuss below.
	Therefore, it is easy to see how, if the interaction coupling $G$ is large enough, the second term wins over the first term and the inverse propagator becomes negative. 
	
	Note that the stability analysis remains blind as to the nature of the instability. Thus, ideally, the two methods can be combined. 
	Once the stability analysis reveals the homogeneous state to be unstable in general one can use the ansatz method to search for 
	specifically favoured inhomogeneous states. 
	
	\subsection{Beyond Models of QCD}
	Within full QCD -- or with any approximated technique that can be systematically improved towards full QCD -- the same two choices remain, direct ansatz or stability analysis. As already noted above, in Ref.~\cite{Muller:2013tya} a chiral density wave-like ansatz was tested for QCD within a DSE framework 
	and a region of the phase diagram was found which allows for self-consistent solutions of the inhomogeneous DSEs with a CDW-like shape.
	
	As for a stability analysis, no such work exists as of yet. Besides QCD being far more complicated to work numerically, this is mostly 
	due to the fact that the current framework described above is not suited for QCD. One cannot write the thermodynamic potential $\Omega$ 
	as a functional of the condensates as for NJL and other models. This is because, in QCD, the self-energy or the inverse propagators 
	are not simple linear combinations of the condensates. In fact, the same problem would appear beyond mean-field in NJL or QM models\footnote{
		This is indeed relevant as there is lively discussion about quantum fluctuations of Nambu-Goldstone bosons related to the symmetry breaking disordering the inhomogeneous system \cite{Lee:2015bva,Hidaka:2015xza,Yoshiike:2017kbx} , e.g., into a Quantum-Pion-Liquid (QPL) or a Liquid-Crystal (LC)~\cite{Pisarski:2020dnx,Pisarski:2021qof,Rennecke:2021ovl}, which further corroborates the importance of developing a method that goes beyond mean field within the scope of a stability analysis.}.
	Therefore, the main goal of this paper is to show an alternative method to perform a stability analysis that both reproduces the 
	classic NJL results and is applicable to QCD, beyond mean-field models, and, in fact, any field theory.

	\section{2PI Effective Action Stability Analysis\label{sec:staba}}
	
	In order to formulate the framework for a general stability analysis, it is first useful to recall some basic facts. In a classical field 
	theory, the system tends towards minima of the classical action $S_{cl}$. The equations of motion of the theory are given by extrema conditions
	$\delta S_{cl} / \delta \phi =0$ of the classical action with respect to the fields. In a quantum field theory, the effective action $\Gamma$ 
	plays the same role and the corresponding quantum equations of motion, the Dyson-Schwinger equations, are found by exactly the same logic. 
	Also, the effective action is a proxy for the pressure and other thermodynamic quantities of the theory. The thermodynamic potential,
	$\Omega$, at finite temperature $T$ and finite chemical potential $\mu$ is found by subtracting the vacuum term\footnote{In order to avoid notation overload, we will use the round bracket notation $\Gamma(T,\mu)$ to signify the functional $\Gamma$ evaluated at the stationary point for fixed temperature and chemical potential, i.e. $\Gamma[S(T,\mu),T]\vert_\text{Stationary}$. Note that $\Gamma$ not only depends on $T$ via the propagator $S$ but also explicitly in the Matsubara sums.}
	\begin{equation}\label{thermorels}
	\Omega(T,\mu) = -\frac{T}{V} \Big(
	\Gamma(T,\mu) - \Gamma(0,0)
	\Big)= -\frac{T}{V} P(T,\mu).
	\end{equation}
	Furthermore, one can write the effective action of a theory as a functional of the n-point functions of the fields, 
	the inverse propagators and vertices (see e.g. the \ref{appendixnpoint} and Refs.~\cite{berges2004n,carrington2015techniques}).

	Starting from a 2-Particle-Irreducible (2PI) effective action \cite{Cornwall:1974vz} such as
	\begin{equation}\label{gamma}
	\Gamma = 
	\text{Tr} \log \left[ {S^{-1}} \right] 
	-\text{Tr}\left[\boldsymbol{1}-S_{0}^{-1} S\right]
	+\Phi_\text{2PI}[S],
	\end{equation}
	where, $S$ is the full propagator, $S_0$ the bare one, and $\Phi_\text{2PI}[S]$ is the sum of all 2PI diagrams, we find the Dyson-Schwinger equations by
	\begin{equation}
	\label{DSE2PI}
	\frac{\delta \Gamma}{\delta S}=0 \quad \Rightarrow\quad 
	-S^{-1} 
	+S_{0}^{-1}
	+{\frac{\delta \Phi_\text{2PI}}{\delta S}}=0
	\quad \Rightarrow\quad 
	S^{-1} 
	=S_{0}^{-1}+\Sigma,
	\end{equation}
	where we identified the derivative of $\Phi_\text{2PI}$ with the self-energy $\Sigma$. Solutions to the equation above 
	will reveal every stable, unstable or metastable equilibrium configuration of the system. Let $\bar{ S}$ be an equilibrium solution, 
	namely the homogeneous configuration. We may analyse if it is stable or not by taking
	$$S = \bar{S} + \delta S.$$
	where $\delta S$ is a small inhomogeneous propagator (what sufficiently ``small'' means is described in a later section).
	We can then take a functional Taylor expansion around the $\delta S =0$ point
	\begin{equation}\label{gamma2expansion}
	\begin{aligned}
	\Gamma[S] &=\Gamma[\bar{S}] 
	+\text{Tr}
	\left[\,\overline{\frac{\delta \Gamma}{\delta S_{xy}}} \delta S_{yx}\right]
	+\frac{1}{2!}\text{Tr}
	\left[\,\overline{\frac{\delta^2 \Gamma}{\delta S_{xy} \delta S_{zs}}}\delta S_{yx} \delta S_{sz}\right]
	+\cdots\end{aligned}
	\end{equation}
	where we again use the bar notation $\bar O$ to denote $O$ evaluated at the homogeneous solution. The indices on the various $\delta S$ will from 
	now on be omitted and taken to be implicitly understood (see \ref{appendixder}). We then have
	\begin{align}
	\Gamma^{(0)}&= - \text{Tr}\log[\bar{S}]
	-\text{Tr}\left[\boldsymbol{1}-S_{0}^{-1}\bar{S}\right]
	+ \Phi_\text{2PI}[\bar{ S}]
	\nonumber
	\\
	\Gamma^{(1)}&=
	\text{Tr}\left[\left(-\bar S^{-1} 
	+S_{0}^{-1}
	+\overline{\frac{\delta \Phi_\text{2PI}}{\delta S}}\right)\delta S\right]
	\nonumber
	\\
	\Gamma^{(2)}&= 
	\frac{1}{2}\text{Tr}[(\bar S^{-1} \delta S)^2]
	+\frac{1}{2}\text{Tr}
	\left[\overline{\frac{\delta^2 \Phi_\text{2PI}}{\delta S \delta S}}\delta S \delta S\right]\label{Gamma_expansion}
	\end{align}
	where $\Gamma^{(0)}$ is the homogeneous effective action, $\Gamma^{(1)}$ vanishes at the stationary 
	point\footnote{This is trivially so, given that the first-order Taylor contribution comes with the first derivative 
		of $\Gamma$ with respect to $S$ which is zero the stationary point by definition.}, and the thermodynamic potential 
	\begin{equation}\label{TDstability}
	\Omega^{(2)}(T,\mu) = -\frac{T}{V} \Big(\Gamma^{(2)}(T,\mu)	\Big)
	\end{equation}
	arising from $\Gamma^{(2)}$ is our starting point for the stability analysis.


	\subsection{Some remarks on the $n$PI Case \label{sec:npi}}
	
	The approach described above is not solely applicable to 2PI effective actions. A brief look at the more general case may be enlightening. 
	Take a 1PI effective action, which is characterised by the Legendre transform of the connected generating functional $W$ in the presence of 
	a one-point source $J$,
	\begin{equation}
	\Gamma[\phi] = W[J] - \phi_i J_i,
	\end{equation}
	where here we use the standard simplified notation of two repeated indices implying sum and integration of all discrete indices and continuum 
	variables respectively.
	As per usual, the root of
	\begin{equation}
	\frac{\delta \Gamma}{\delta \phi_i} = -J_i,
	\end{equation}
	defines the physical point, and our one-point function -- most generally, in the presence of sources -- is
	\begin{equation}
	\frac{\delta W}{\delta J_i} = \phi_i.
	\end{equation}
	The second derivatives yield the following relations
	\begin{equation}
	\frac{\delta^2 \Gamma}{\delta \phi_i\delta \phi_j}=-\frac{\delta J_i}{\delta \phi_i},
	\qquad
	\frac{\delta^2 W}{\delta J_i\delta J_j} = \frac{\delta \phi_i}{\delta J_j}.
	\end{equation}
	Assuming, as per usual, invertibility of $\phi[J]$, we obtain
	\begin{equation}\label{basicG2}
	\frac{\delta^2 \Gamma}{\delta \phi_i \delta \phi_j} = -\left(\frac{\delta^2 W}{\delta J_j\delta J_i}\right)^{-1}.
	\end{equation}
	It is an elementary result in quantum field theory that $n$ derivatives of the generating functional $W$ with respect to the one-point sources $J$ yield the connected $n$-point functions (see \ref{appendixnpoint}). Therefore, Eq.~(\ref{basicG2}) shows that the second derivative of the 1PI effective action with respect to the 1-point function is (minus) the inverse 2-point function\footnote{Since the calculation is based on a 1PI effective action the 2-point function is \textit{not} defined self-consistently.} in the presence of sources.
	
	Note, however, that the NJL result in Eq.~(\ref{classicNJLcondition}) \textit{is} based on a 1PI approach. This comes from the fact that the model is written as a field theory of the mean-field condensates (the  $\Omega^\text{NJL}_\text{MF}[\phi]$ defined above).
	The question is, can this approach be generalised? 
	Indeed, take a 2PI effective action, i.e., a functional of two functions whose stationary point defines \textit{self-consistently} the one-point and two-point functions
	\begin{equation}
	\Gamma[\phi, S] = W[J,R] - \phi_iJ_i - S_{ij}R_{ij}.
	\end{equation}
	The same game can be played and we obtain
	\begin{equation}
	\frac{\delta^2 \Gamma}{\delta \phi_i\delta \phi_j}=-\frac{\delta J_i}{\delta \phi_i},
	\qquad
	\frac{\delta^2 W}{\delta J_i\delta J_j} = \frac{\delta \phi_i}{\delta J_j},
	\qquad
	\frac{\delta^2 \Gamma}{\delta \phi_i \delta \phi_j} = -\left(\frac{\delta^2 W}{\delta J_j\delta J_i}\right)^{-1},
	\end{equation}
	\begin{equation}
	\frac{\delta^2 \Gamma}{\delta S_{ij}\delta S_{kl}}=-\frac{\delta R_{ij}}{\delta S_{kl}},
	\qquad
	\frac{\delta^2 W}{\delta R_{ij}\delta R_{kl}} = \frac{\delta S_{ij}}{\delta R_{kl}},
	\qquad
	\frac{\delta^2 \Gamma}{\delta S_{ij} \delta S_{kl}} = -\left(\frac{\delta^2 W}{\delta R_{kl}\delta R_{ij}}\right)^{-1}
	\end{equation}
	The second derivative of $W$ with respect to the two-point source $R$ is the \textit{inverse} four-point function.
	If we now revisit Eq.~(\ref{gamma2expansion}) we see that, at the physical point, it is trivial that $\Gamma^{(1)}$ always vanishes and $\Gamma^{(2)}$ is given by (minus) the inverse 2-point function. Therefore, if we assume both the one- and two-point functions can be inhomogeneous, our instability condition will be, analogously to Eq.~(\ref{gamma2expansion}), given by
	\begin{equation}
	\Gamma^{(2)}=\frac{1}{2!}\text{Tr}
	\left[\,\overline{\frac{\delta^2 \Gamma}{\delta \phi_x \delta \phi_y}}\delta \phi_x \delta \phi_y\right]
	+
	\frac{1}{2!}\text{Tr}
	\left[\,\overline{\frac{\delta^2 \Gamma}{\delta S_{xy} \delta S_{zs}}}\delta S_{yx} \delta S_{sz}\right]
	\end{equation}
	and thus, negativity of the connected two-point function or the connected four-point function will signal an instability.
	In QCD we do not have an analogous to the one-point function $\phi$. Therefore, an instability will be signalled by negativity of the four-point functions of the theory.
	
	It is trivial to see how this can be generalised to any $m$-loop $n$PI effective action and the {in}stability condition will always be given by negativity of the inverse $2n$-point functions. More explicitly, take an $n$PI effective action
	$$
	\Gamma[\phi,S_2,S_3,S_4,\dots,S_n],
	$$
	which at the stationary point gives, self-consistently, the fully-dressed $n$-point functions of the theory. Negativity of the second derivative of $\Gamma$ with respect to any $\phi,S_2,S_3,S_4,\dots,S_n$ at the stationary point -- or, in other terms, negativity of any of the inverse two-particle, four-particle, $\dots$, $2n$-particle connected Green's functions -- is the {in}stability condition.


	\section{Application to the NJL model}
	\label{sec:NJL}
	
	As a first test, we apply the method developed in section \ref{sec:staba} to the NJL model, \Eq{LNJL}.
	In Hartree approximation, which was also the basis of \Eqs{classicNJLcondition} and (\ref{DM}),
	we have the 2PI functional
	\begin{equation}
	\Phi_\text{2PI} = G \int_{x,y} \delta(x-y) \sum_M
	\text{tr}\big[V^{(M)}S(x,x)\big]
	\times
	\text{tr}\big[V^{(M)}S(y,y)\big]
	\end{equation}
	with the vertices  $V^{(M)}$ defined below \Eq{LNJL}
	and the integrals are taken over the Euclidean four-volume.
	As before  the, lower case ``tr'' notation means a trace over internal degrees of freedom.
	From \Eq{DSE2PI} we then obtain the quark self-energy
	\begin{equation}\label{localsigma}
	\Sigma(x,y) = {\frac{\delta \Phi_\text{2PI}}{\delta S(y,x)}} = 2G\, \delta(x-y) \sum_M \text{tr}\big[V^{(M)}S(x,x)\big] V^{(M)}
	\equiv \Sigma(x) \delta(x-y),
	\end{equation}
	making explicit that the self-energy is local, i.e., depends only on a single space-time coordinate. 
	Of course, this is an expected consequence of the local four-point interaction together with the Hartree approximation, but it will become
	important later on. 
	
	In homogeneous ($=$ translationary invariant) matter the propagator depends only on space-time differences,
	$\bar{S} = \bar{S}(x-y)$, so that the self-energy  becomes constant:
	\begin{equation}
	\bar{\Sigma}(x) = 2G\, \sum_M \text{tr}\big[V^{(M)}\bar{S}(0)\big] V^{(M)} = \mathit{const.}
	\end{equation}
	In the following, we will refer to this equation as the homogeneous gap equation.
	
	As outlined in the previous section, we now consider small, possibly inhomogeneous,  fluctuations $\delta S(x,y)$ of the propagator around the homogeneous solution
	$\bar S$, and expand the effective action in powers of $\delta S$. At first order we obtain
	\begin{equation}\label{Gamma1NJLconfig}
	\Gamma^{(1)} 
	= -\int_{x,y} \text{tr}\big[\bar{\Sigma}(x,y) \delta S(x,y)\big]
	+  2G\,\sum_M \int_x  \text{tr}\big[V^{(M)}\bar{S}(0)\big] \text{tr}\big[V^{(M)}\delta S(x,x)\big], 
	\end{equation}
	where the first term on the right-hand side originates from the two functional traces in \Eq{gamma} and the last term is due to
	$\Phi_\text{2PI}$. Using the locality of the self-energy together with the homogeneous gap equation, we immediately see
	that $\Gamma^{(1)} = 0$, confirming our general expectation.
	
	At second order we have
	\begin{equation}
	\Gamma^{(2)} 
	= \frac{1}{2} \int_{w,x,y,z} \text{tr} \big[\bar{S}^{-1}(w,x) \delta S(x,y) \bar{S}^{-1}(y,z) \delta S(z,w)\big]
	+
	G\,\sum_M \int_x
	\left(\text{tr}\big[V^{(M)}\delta S(x,x)\big]\right)^2,
	\end{equation}
	where we explicitly wrote the functional trace in coordinate space. Turning to momentum space this becomes
	\begin{equation}
	\label{Gamma2NJLmom}
	\begin{aligned}
	\Gamma^{(2)} 
	=  &\frac{1}{2} \SumInt_{p,q} \text{tr} \big[\bar{S}^{-1}(p+q) \delta S(p+q,p) \bar{S}^{-1}(p) \delta S(p,p+q)\big]
	\\
	&+
	G \sum_M \SumInt_q 
	\Big(\SumInt_{p} \text{tr}\big[V^{(M)}\delta S(p+q,p)\big] \Big)\Big(  \SumInt_{k} \ \text{tr}\big[V^{(M)}\delta S(k,k+q)\big]\Big).
	\end{aligned}
	\end{equation}
	As discussed, a positive $\Gamma^{(2)}$, corresponding to a negative contribution to the thermodynamic potential, signals an 
	instability of the homogeneous solution. However, if we compare the above expression with \Eq{classicNJLcondition}, we see
	that in the latter the moduli squared of the fluctuations appear as isolated factors, so that the question about the stability of the
	homogeneous state could be decided solely on the sign of their coefficients $D_M^{-1}$. 
	In \Eq{Gamma2NJLmom}, on the other hand, isolating the fluctuations from the rest is inhibited by the different momentum arguments.
	As a consequence, a modulation-shape agnostic stability analysis is not immediately possible on the basis of this equation.
	
	However, we have not yet exploited the fact that the quark self-energy and, hence, its fluctuations $\delta\Sigma$ are local.
	To this end it is more appropriate to expand the effective action in powers of $\delta\Sigma$, corresponding to fluctuations 
	of the inverse propagator, rather than in fluctuations of the propagator itself. This can be achieved
	by using the Dyson series,
	\begin{equation}
	S^{-1} = \bar{S}^{-1} + \delta\Sigma
	\;\Leftrightarrow\;
	S = \bar{S} - \bar{S}\delta\Sigma S =  \bar{S} - \bar{S}\delta\Sigma \bar{S} + \bar{S}\delta\Sigma \bar{S} \delta\Sigma \bar{S}
	- \dots = \bar{S} + \delta S,
	\end{equation}
	and inserting the resulting $\delta S$ into the above expressions up to the desired order in $\delta\Sigma$.  
	Since $\Gamma^{(1)}$ vanishes to all orders by the gap equation, we immediately see that 
	in a $\delta\Sigma$  expansion of the effective action the first-order term vanishes as well.
	Moreover, the order $(\delta\Sigma)^2$ contributions can only arise from inserting the order $\delta\Sigma$ 
	propagator fluctuations $\delta S^{(1)} = - \bar{S}\delta\Sigma \bar{S}$
	into $\Gamma^{(2)}$.
	In momentum space, as a consequence of the locality,  $\delta\Sigma$ depends only on a single momentum variable,
	and $\delta S^{(1)}$ is given by
	\begin{equation}
	\delta S^{(1)}(k_1,k_2) = -\bar S(k_1) \delta\Sigma(k_1-k_2) \bar S(k_2).
	\end{equation}
	Inserting this into \Eq{Gamma2NJLmom} we find
	\begin{equation}
	\begin{aligned}
	\tilde\Gamma^{(2)} 
	=  &\frac{1}{2} \SumInt_{p,q} \text{tr} \big[\delta\Sigma(q) \bar{S}(p) \delta\Sigma(-q) \bar{S}(p+q)\big]
	\\
	&+
	G \sum_M \SumInt_q 
	\Big(\SumInt_{p} \text{tr}\big[V^{(M)}\bar S(p+q) \delta\Sigma(q) \bar S(p) \big] \Big)
	\Big(\SumInt_{k} \text{tr}\big[V^{(M)}\bar S(k) \delta\Sigma(-q) \bar S(k+q) \big] \Big),
	\end{aligned}
	\end{equation}
	where the tilde in $\tilde\Gamma^{(2)}$ was introduced to indicate that we have now performed an expansion in $\delta\Sigma$ rather than in $\delta S$.
	
	In order to proceed we decompose the fluctuations into scalar and pseudoscalar parts,
	\begin{equation}
	\delta\Sigma(q) = \sum_M \delta\Sigma_M(q) V^{(M)}, 
	\end{equation}
	and pull the functions $\delta\Sigma_M$ out of the traces.
	The latter vanish for  unequal vertex factors $V^{(M)} \neq V^{(M')}$, so we are left with a single sum over $M$ for both terms.
	Moreover, for local fluctuations $\delta\Sigma_M$ are real functions in coordinate space.
	In momentum space they therefore satisfy the relation $\delta\Sigma_M(-q) = \delta\Sigma^*_M(q)$
	and, hence, their product is given by $|\delta\Sigma_M(q)|^2$.
	Combining all terms $\tilde\Gamma^{(2)}$ can then be written as
	\begin{equation}\label{njlGamma2}
	\begin{aligned}
	\tilde\Gamma^{(2)} 
	=  &G  \sum_M \SumInt_{q}  |\delta\Sigma_M(q)|^2
	\SumInt_{p} \text{tr} \big[V^{(M)} \bar{S}(p) V^{(M)}  \bar{S}(p+q)\big]
	\underbrace{\left(\frac{1}{2G} + \SumInt_{k} \text{tr}\big[V^{(M)}\bar S(k) V^{(M)}  \bar S(k+q) \big] \right)}_{D_M^{-1}(q)},
	\end{aligned}
	\end{equation}
	where we identified the inverse meson propagators from \Eq{DM}.
	
	Again, since we assume the self-energy fluctuations to be time independent, their Fourier modes are restricted to vanishing energy,
	$\delta\Sigma(q) = \frac{1}{T} \delta_{q_4,0} \, \delta\Sigma(\vq)$ see \ref{appendixFourier}.
	From \Eq{TDstability} we then obtain for the second-order contribution of the thermodynamic potential
	\begin{equation}
	\label{tildeOmega2}
	\tilde\Omega^{(2)} = - \frac{T}{V}\tilde\Gamma^{(2)}
	=
	-\frac{G}{V}  \sum_M \int\limits_{\vq}  |\delta\Sigma_M(\vq)|^2 \,D_M^{-1}(q) \, \ell_M(q),
	\end{equation}
	with the loop integral
	\begin{equation}
	\ell_M(q) = \SumInt_{p} \text{tr} \big[V^{(M)} \bar{S}(p) V^{(M)}  \bar{S}(p+q)\big].
	\end{equation}
	Recall that an instability of the homogeneous ground state against small fluctuations corresponds to a negative $\Omega^{(2)}$. 
	However, if we compare \Eq{tildeOmega2} with the literature NJL result  \Eq{classicNJLcondition}
	an essential difference seems to be the additional factor $-\ell_M(q)$ in the integrand of the former.
	In order to get a deeper understanding of the role of this factor it is instructive to consider the limit of vanishing interactions,
	i.e., a free fermi gas. For this case $\tilde\Omega^{(2)}$ becomes
	\begin{equation}
	\label{tildeOmega2ffg}
	\tilde\Omega^{(2)}_\text{ffg} = \lim\limits_{G\rightarrow 0} \tilde\Omega^{(2)} 
	=
	-\frac{1}{2V}  \sum_M \int\limits_{\vq}  |\delta\Sigma_M(\vq)|^2  \, \ell_M(q),
	\end{equation}
	showing that the stability of the free fermi gas is determined by $\ell_M(q)$. 
	More precisely, since we expect the free fermi gas to be stable against inhomogeneous fluctuations, we can conclude that
	$\ell_M(q)$ is always negative for $q = (\vq, q_4 = 0)$. This physical argument is backed up by numerical calculations.
	
	Returning to \Eq{tildeOmega2}, the negativity of $\ell_M(q)$ means that in the interacting case the (in)stability of the homogeneous
	ground state is entirely determined by the sign of the inverse meson propagators.
	Exactly as in \Eq{classicNJLcondition} $D_M^{-1}(q) < 0$ in some momentum regime indicates an instability, while the absence of such
	negative regions means that the homogeneous ground state is stable against small inhomogeneous fluctuations.
	In this context it is interesting to note that $D_M^{-1} = \frac{1}{2G} + \ell_M$. Thus, the negativity of $\ell_M$ which guarantees the 
	stability of the non-interacting fermi gas, at the same time opens the possibility for an instability in the interacting system.
	
	So, in summary, we managed to recover the standard result, \Eq{classicNJLcondition}, within the 2PI approach. 
	To this end, however, it was crucial to exploit the locality of the quark self-energy, which enabled us to isolate a factor 
	$ |\delta\Sigma_M(\vq)|^2$ and in turn allowed for a modulation-shape agnostic stability analysis.  
	Without utilizing the locality this was not possible, as can be seen from \Eq{Gamma2NJLmom}
	where the different momentum arguments of the propagator fluctuations inhibit that they can be pulled out of the innermost integrals
	and be combined to an absolute value squared. 
	As we will see in the next section, we will encounter the same difficulty in QCD where the gluon-mediated interaction leads 
	to genuine non-local quark self-energies.
	
	
	\section{General considerations for QCD applications of the stability analysis \label{sec:fullsa}}
	The application of the method described above is far more intricate for QCD than for the NJL-model and a brief discussion 
	of such difficulties is enlightening. Recall, that our general starting point for the stability analysis is the expression 
	Eq.~(\ref{TDstability}) for the second order expansion of the thermodynamic potential. In the mean field NJL model the 
	corresponding expressions Eq.~(\ref{classicNJLcondition}) and Eq.~(\ref{tildeOmega2}) both feature the property that the 
	inhomogeneous contribution be factorised into a ``mod-squared'' term such as $|\delta \Sigma(q)|^2$ in Eq.~(\ref{tildeOmega2}) or 
	$|\delta \phi_{M}(q)|^2$ in Eq.~(\ref{classicNJLcondition}). This renders a stability analysis particularly 
	simple, since no assumptions on the specific form of the instability have to be made.
	
	However, in QCD,
	even with the drastic approximation of bare quark-gluon vertices, due to the non-local interaction term, 
	it will not be like this. Take the following 2PI effective potential
	\footnote{Dots denote dressed quantities;
		solid and wiggly lines represent quarks and gluons, respectively. All vertices are here bare.}
	\begin{equation}
	 \Phi_\text{2PI}[S]=\frac{1}{2}
	 \raisebox{-0.8cm}{
	\begin{tikzpicture}
	\begin{feynman}
	\vertex (a);
	\vertex [right=of a] (b);
	\vertex [right=of a] (b);
	\vertex [right=0.75cm of a] (l);
	\diagram*{
		(a) -- [fermion3,half left] (b);
		(b) -- [boson] (a);
		(b) -- [fermion3,half left] (a);
	};\draw (l) node [dot];
	\end{feynman}
	\end{tikzpicture}
	}
	=\frac{1}{2}\text{Tr}
	\left[
	\gamma_\mu t^a  S(k_1,k_2) \gamma_\nu t^b
	S(k_2-q,k_1-q) D_{\mu\nu}^{ab}(q)
	\right]
	\end{equation}
	where $D_{\mu\nu}^{ab}$ is the gluon propagator. 
	From Eq.~(\ref{Gamma_expansion}) we get
	\begin{equation}
	\text{Tr}
	\left[\overline{\frac{\delta^2 \Phi_\text{2PI}}{\delta S \delta S}}\delta S \delta S\right]=
	\text{Tr}
	\left[
	\gamma_\mu t^a \delta S(k_1,k_2) \gamma_\nu t^b
	\delta S(k_2-q,k_1-q) D_{\mu\nu}^{ab}(q)
	\right],
	\end{equation}
	in which case the $\delta S(k_1,k_2)$ and $\delta S(k_2-q,k_1-q)$ terms {\it cannot} be simply factored out into 
	a $|\delta S|^2$-type multiplicative factor. Thus the specific form of these terms do play a role. At some point or another in the analysis, we will have to specify what $\delta S$ is. This  
	brings about a few complications which need to be incorporated into the analysis, as will be discussed in the following. 
	
	Conceptually, the analysis described above is very simple. It can be framed in the following way. First, we find stationary points of the effective action (denoted by the bar notation). Then we perform the Taylor expansion which, at the end of the day, results in calculating a second derivative
	\begin{equation}\label{eq:2ndDeriv}
	\frac{1}{2!}\text{Tr}
	\left[\,\overline{\frac{\delta^2 \Gamma}{\delta S_{xy} \delta S_{zs}}} \delta S_{yx}  \delta S_{sz}\right].
	\end{equation}
	If it shows positive/zero/negative curvature, it will indicate the stationary point is a maximum/saddle/minimum.
	This is a \textit{directional derivative} in the infinite-dimensional functional space of which the propagator is an element.
	The natural analog with $\mathbb{R}^N$ case is
	$$
	\text{first:}\quad \vec{u}\cdot\frac{\partial f(\vx)}{\partial \vx},\qquad
	\text{second:}\quad \vec{u}\cdot\frac{\partial }{\partial \vx}\left(\vec{u}\cdot\frac{\partial f(\vx)}{\partial \vx}\right),
	$$
	i.e. the derivative of $f$ in the $\vec{u}$ direction, corresponds to
	$$
	\text{first:}\quad\Tr\left[\frac{\delta F[\phi]}{\delta \phi(x)} \lambda(x)\right],\qquad
	\text{second:}\quad\Tr\left[\frac{\delta^2 F[\phi]}{\delta \phi(x)\delta \phi(y)} \lambda(y)\lambda(x)\right],
	$$
	the derivative of $F$ in the direction of some test-function $\lambda(x)$. This is sometimes denoted in the literature as $\mathcal{D}_{\lambda(x)}F[\phi(x)]$. What we are calculating, then, is the directional derivative of the effective action 
	in the direction of $\delta S = S-\bar{S}$. In order to make this a well-defined process the test-function needs to satisfy 
	at least the following properties:
	\begin{itemize}
		\item It is a function $\delta S:\mathbb{R}^n\rightarrow\mathbb{C}$ which must go to zero at infinity.
		\item It must be a ``small'' contribution to the propagator. This is rather subtle, however, it is sufficient to conceive of a sequence $\delta S_\lambda$ such that $\lim_{\lambda \rightarrow 0} \delta S_\lambda = 0$ {uniformly}. A ``small'' $\delta S$ is then taken to be understood as a large $\lambda$ case of $\delta S_\lambda$. If $\delta S(k)_\lambda = \lambda f(k)$, where $f(k)$ is limited, this is sufficient. Moreover, the $\lambda$ factors out of $\Gamma^{(2)}$ and one can ignore it. 
		\item It must follow the propagator's adjoint relation
		\begin{equation}\label{adjoint}
		\delta S(\omega_1,\vk_1,\omega_2,\vk_2)^\dagger = \gamma_4 \delta S(-\omega_2,\vk_2,-\omega_1,\vk_1) \gamma_4 \,,
		\end{equation}
		without which the pressure can turn out complex.
	\end{itemize}
	Taking this derivative in the direction of a test-function which does not follow these properties will lead to incorrect results.
	Most importantly, though, the imaginary part of the test-function must be constrained. This is because $\Omega^{(2)}[\delta S]$ is not limited from below and every stationary point is a saddle point with respect to the imaginary part of the propagator. This is easily verified, since $\Omega^{(2)}$ is quadratic in the $\delta S$ function. Let $\delta S_1$ be a real function which follows the properties listed above and such that $\Omega^{(2)}[\delta S_1]>0$. Take $\delta S_2 = i\alpha \delta S_1$. Naturally $\Omega^{(2)}[\delta S_2] = -\alpha^2 \Omega^{(2)}[\delta S_1]$ so, the larger $\alpha$, the smaller the free energy. Meaning that the free energy is unlimited from below with respect to the imaginary part of $\delta S$. We chose then to constrain the imaginary part of the test-function by the following condition:
	\begin{equation}\label{eq:MBFconjecture}
	\Tr\left[ \frac{\delta \Omega^{(2)}}{\delta S} \text{Im}(\delta S) \right] =0.
	\end{equation}
	The condition in Eq.~(\ref{eq:MBFconjecture}) fundamentally maximises $\Omega^{(2)}$ with respect to the imaginary part of $\delta S$ and thus, effectively protects the analysis against false instabilities that could appear due to an inadequate $\delta S$. Here thereafter, Eq.~(\ref{eq:MBFconjecture}) will be understood as an integral part of the analysis.

	\section{Chiral Phase Transition Test \label{sec:chiral}}
	The approach described in section~\ref{sec:staba} and \ref{sec:fullsa} is not \textit{only} valid for inhomogeneous phases. Whatever the nature of the phase 
	transition may be, it can be tested via stability analysis as long as the energetically unfavourable phase sits either on 
	a maximum or a saddle point of the thermodynamic potential. This is the case, for instance, for the chirally symmetric phase of massless QCD at the chiral second order phase boundary to the {\it homogeneous} broken phase. In this section we therefore want to test the method outlined above by applying it to study that phase transition.
	
	In the chiral limit with vanishing explicit quark masses, QCD has a chiral-symmetric solution 
	for all temperatures and densities, i.e. a massless propagator of the form
	\begin{equation}
	S^{-1}(k) = i \slashed{\vk}A(k) +i(k_4+i\mu)\gamma_4 C(k),
	\end{equation}
	together with a broken, massive, solution of the form
	\begin{equation}
	S^{-1}(k) = i \slashed{\vk}A(k) +B(k) +i(k_4+i\mu)\gamma_4C(k),
	\end{equation}
	where A, B, and C are so-called dressing functions which
	emerge as a self-consistent solutions of the quark Dyson-Schwinger equation. In the high temperature/chemical potential phase, this solution 
	is a minimum of the thermodynamic potential, whereas in the hadronic low temperature/chemical potential phase it is at a maximum 
	and therefore unstable. We can analyse the (in)stability of such a solution by the method above, assuming the propagator is of the form
	\begin{equation}\label{deltaSchiral}
	\begin{aligned}
	S(k)
	&= \frac{-i \slashed{\vk}A(k) +\delta m(k)-i(k_4+i\mu)\gamma_4 C(k)}{\vk^2A(k)^2+(k_4+i\mu)^2C(k)^2}=
	\underbrace{\frac{-i \slashed{\vk}A(k)  -i(k_4+i\mu)\gamma_4 C(k)}{\vk^2A(k)^2+(k_4+i\mu)^2C(k)^2}}_{\bar{S}~\text{chiral}}
	+
	\underbrace{\frac{ \delta m(k)}{\vk^2A(k)^2+(k_4+i\mu)^2C(k)^2}}_{\delta S~\text{breaks chiral symmetry}}
	\end{aligned}
	\end{equation}
	that is, a chiral propagator plus a \textit{small} massive term which breaks chiral symmetry ($\delta m(k)$ is effectively a small ``$B$'' function). Using this form we may calculate 
	$\Gamma^{(2)}$ explicitly. Note that in this case (chiral instability) we explicitly know the form of the instability (Eq.~(\ref{deltaSchiral})) and what 
	is left is to explore different forms of the mass function $\delta m(k)$ in our test-function.
	
	\subsection{Truncation}
	In order to be able to carry out the explicit numerical calculations, we need to specify the truncation of QCD. 
	The exact DSE for the quark propagator is shown diagrammatically in Fig.~\ref{fig:qDSE}.
	
	\begin{figure}[t]
		\centering%
		\begin{tikzpicture}
		\begin{feynman}
		\vertex (a);
		\vertex [right=of a] (c);
		\vertex [right=0.5cm of c] (d);
		\vertex [above=0.4cm of c] (m1);
		\diagram*{
			(a) -- [fermion3] (c)
		};
		\draw (d) node { \({=}\)};
		\draw (m1) node { \(^{-1}\)};
		\vertex [right=0.5cm of d] (a1);
		\vertex [right=of a1] (c1);
		\vertex [right=0.5cm of c1] (d1);
		\vertex [above=0.4cm of c1] (m12);
		\diagram*{
			(a1) -- [fermion2] (c1)
		};
		\draw (d1) node { \({+}\)};
		\draw (m12) node { \(^{-1}\)};
		\vertex [right=0.5cm of d1] (a2);
		\vertex [right=0.5cm of a2] (gl1);
		\vertex [right=1cm of a2] (b2);
		\vertex [above=0.4cm of b2] (gldot);
		\vertex [right=1cm of b2] (c2);
		\vertex [left=0.5cm of c2] (gl2);
		\diagram*{
			(a2) -- [fermion2] (b2) -- [fermion2] (c2);
			(gl1) -- [boson,half left] (gl2)
		};
		\draw (gldot) node [dot];
		\draw (b2) node [dot];
		\draw (gl2) node [dot];
		\end{feynman}
		\end{tikzpicture}
		\vspace*{-0.5em}%
		\caption{\label{fig:qDSE}%
			The DSE for the quark propagator. Note that two factors of $i$ stemming from the Euclidean quark-gluon vertices have been included already in the overall sign
			of the self-energy diagram and are not part of our definition of the self energy $\Sigma$. 
		}
	\end{figure}
	In explicit form the DSE equation in homogeneous matter reads:
	\begin{equation}
	\left[S(k)\right]^{-1} = Z_{2}\left[S_0(k)\right]^{-1} 
	+ C_{F}\,Z_{1F} \, g^2 \,\SumInt_q\, 
	\gamma_\mu \,S(q)\, \Gamma_\nu(k,q;l)\, D_{\mu\nu}(l), \label{DSEs-1} 
	\end{equation}
	where the momentum routing is $l=(k-q)$, 
	$C_F=\frac{N_C^2-1}{2N_C}$ is the Casimir operator with $N_c=3$, and all $Z$s are renormalisation constants. The (inverse) dressed quark 
	propagator has been discussed above, its bare counterpart is given by
	\begin{equation}
	\left[S_0(k)\right]^{-1} = i \gamma \cdot k + Z_m m\,,
	\end{equation}
	and contains the renormalized quark mass $m$ from the Lagrangian of QCD. Since we work in the chiral limit, $m=0$.
	
	For this proof of concept we use a simple rainbow-ladder type approach where the dressed quark-gluon vertex 
	$\Gamma_\nu(k,q;l)$ is replaced by its bare counterpart, i.e. 
	\begin{equation}
	\Gamma_\nu(k,q;l) = Z_{1F} \,\gamma_{\nu}\,.
	\end{equation}
	In Landau gauge, the gluon propagator is then given by 
	\begin{align}\label{eq:qProp}
	D_{\mu\nu}(l) &= P_{\mu\nu}^{T}(l) D_{T}(l) + P_{\mu\nu}^{L}(l)D_{L}(l)\,,
	\end{align}
	with momentum $l=(\vec{l},\omega_l)$. The bosonic Matsubara frequencies are $\omega_l=\pi T \, 2n_l$.
	The projectors $P_{\mu\nu}^{{T},{L}}$ are transverse (${T}$) and longitudinal (${L}$) with respect
	to the heat bath vector aligned in four-direction and given by
	\begin{align}\label{eq:projTL}
	P_{\mu\nu}^{T} &= \left(1-\delta_{\mu 4}\right)\left(1-\delta_{\nu 4}\right)\left(\delta_{\mu\nu}-\frac{p_\mu p_\nu}{\vec{p}^{\,2}}\right), \hspace*{2cm}
	P_{\mu\nu}^{L} = P_{\mu\nu} - P_{\mu\nu}^{T} \,,
	\end{align}
	where $P_{\mu\nu} = \delta_{\mu\nu} - p_\mu\, p_\nu/p^2$ is the covariant transverse projector.

	In our simple truncation scheme we replace the gluon propagator functions by the temperature and chemical potential 
	independent effective running coupling
	\begin{equation}\label{watson}
	D_{T}(l) = 	D_{L}(l) = \frac{\left(Z_2\right)^2}{g^2 \left(Z_{1F}\right)^2} \frac{8 \pi^2}{\omega^4} D e^{-l^2/\omega^2} \equiv D(l)
	\end{equation}
	with parameters $D=1$GeV$^2$ and $\omega=0.6$GeV \cite{Alkofer:2002bp} and specific powers of renormalization factors
	to preserve multiplicative renormalisability of the DSE. Essentially, the effective running coupling can 
	be viewed as a convolution of the non-perturbative dressing of the gluon and the dressing for the leading $\gamma_\mu$ 
	structure of the quark-gluon vertex neglecting all medium effects. 
	As reviewed in \cite{Fischer:2018sdj}, this simple class of toy-models already displays the 
	essential structure of the QCD phase diagram without, however, any claims of quantitative accuracy.\footnote{Often 
		such an ansatz is supplemented by an ultraviolet log-tail with a specific form 
		dictated by perturbation theory. Here, we omit this log-tail for simplicity. We checked explicitly that all results 
		remain qualitatively unchanged if the log-tail is included. Technically, including the log-tail and therefore establishing
		the correct perturbative ultraviolet behaviour of QCD entails an additional complication: $\Gamma[S]$ is highly divergent 
		even in fully renormalised theories such as QCD and the then necessary subtraction has to be carried out with great numerical 
		precision. This is possible but numerically demanding and we therefore resorted to the pure Gaussian for all calculations
		presented in this work deferring a more refined treatment to a future publication.}.
	
\begin{figure}[t]
		\centering
		\includegraphics[width=0.47\linewidth]{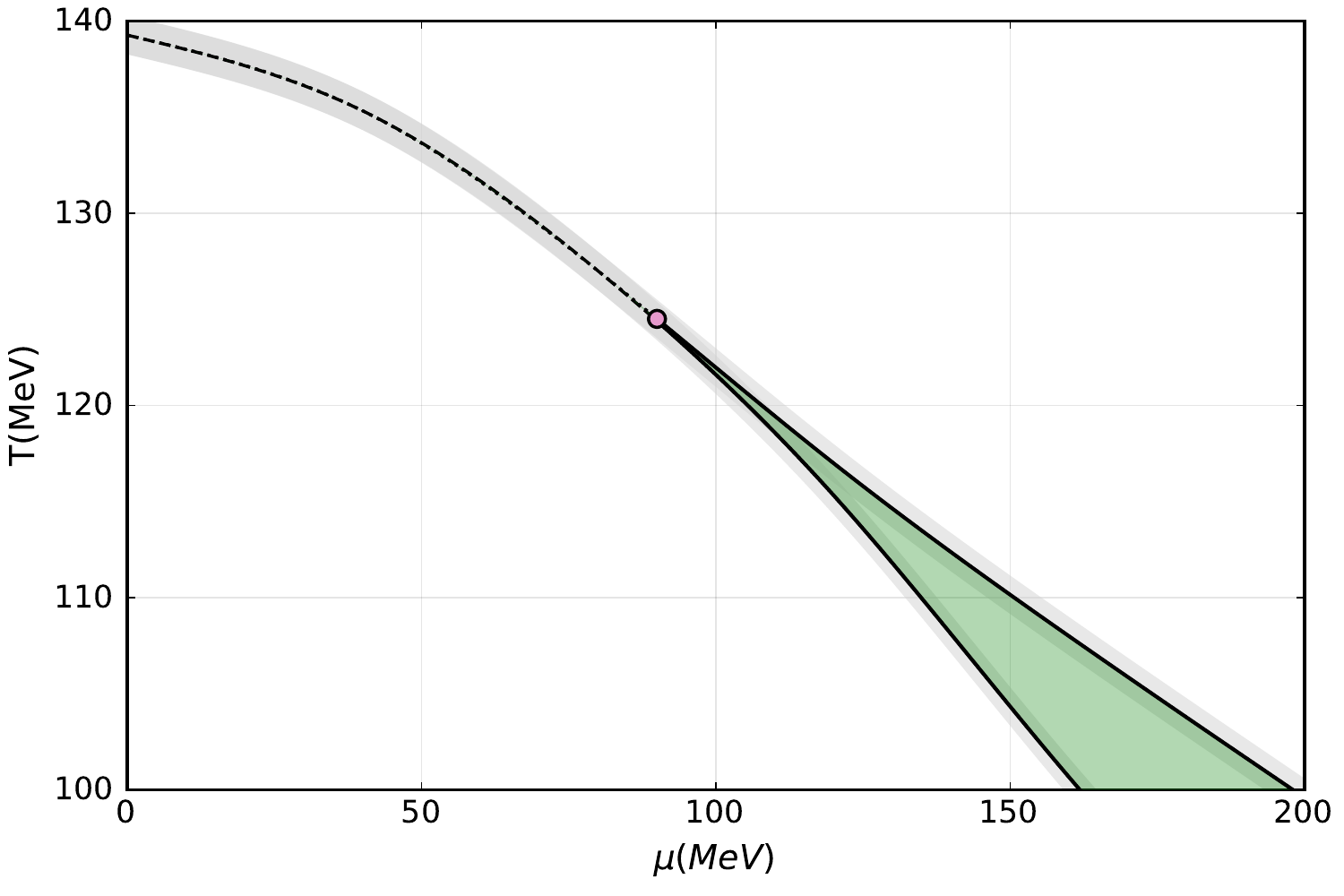}\hfill
		\includegraphics[width=0.47\linewidth]{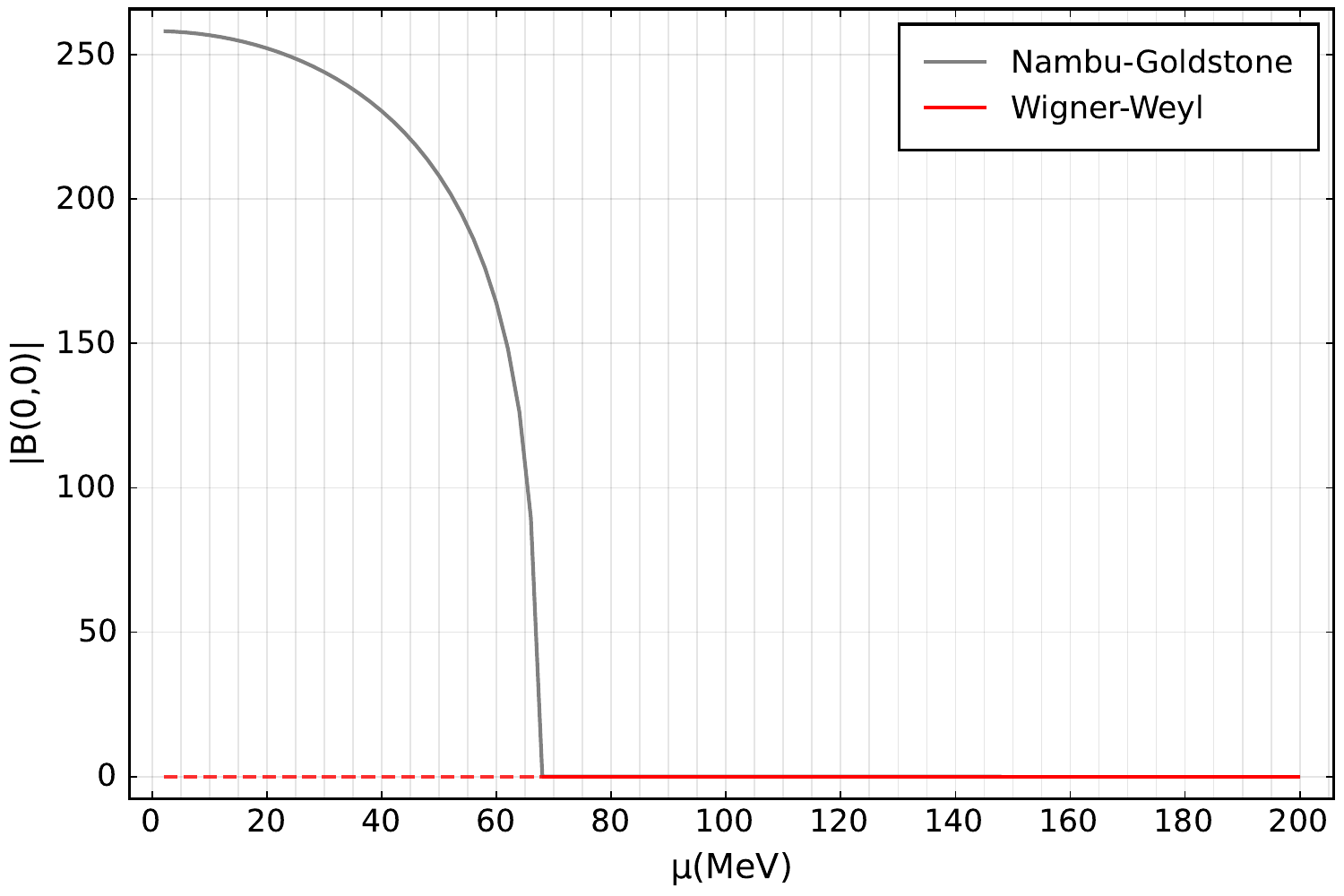}
		\caption{Left diagram: phase diagram for the simple model described in the main text, systematic error of $\pm 1$MeV shown as the grey ribbon. Right diagram: Order parameters for the two solutions at fixed temperature slice $T=130$MeV. The broken, and always stable stable, Nambu-Goldstone solution and the chiral Wigner-Weyl solution.}
		\label{fig:pdslice}
	\end{figure}

	The order parameter for the chiral transition with temperature and chemical potential is the chiral condensate given by 
	\begin{equation} \label{eq:condensate}
	\langle\bar{\psi}\psi\rangle = 
	-Z_2 Z_m  N_c  T\sum_n\int\frac{d^3k}{(2\pi)^3}\mathrm{Tr}_D\left[S(k)\right]\,,
	\end{equation}
	where the trace is taken in Dirac space. Equivalently, also the value of the scalar quark dressing function 
	$B(\vec{k}=0,n_k=0)$ at lowest Matsubara frequency and vanishing three momentum may be used.
	The resulting phase diagram in this model is displayed in the left diagram of Fig.~\ref{fig:pdslice}. One finds the second order 
	chiral transition for small quark chemical potential $\mu$, a critical end-point and a first order transition at large $\mu$.
	At a fixed temperature slice below the critical one at $\mu=0$ we generically find two solutions for the DSEs 
	for small chemical potential up to the transition line. The right hand diagram of Fig.~\ref{fig:pdslice} shows 
	our order parameter for chiral symmetry for both solutions in the region with second order transition.
	\begin{figure}[h!]
		\centering
		\includegraphics[width=1\linewidth]{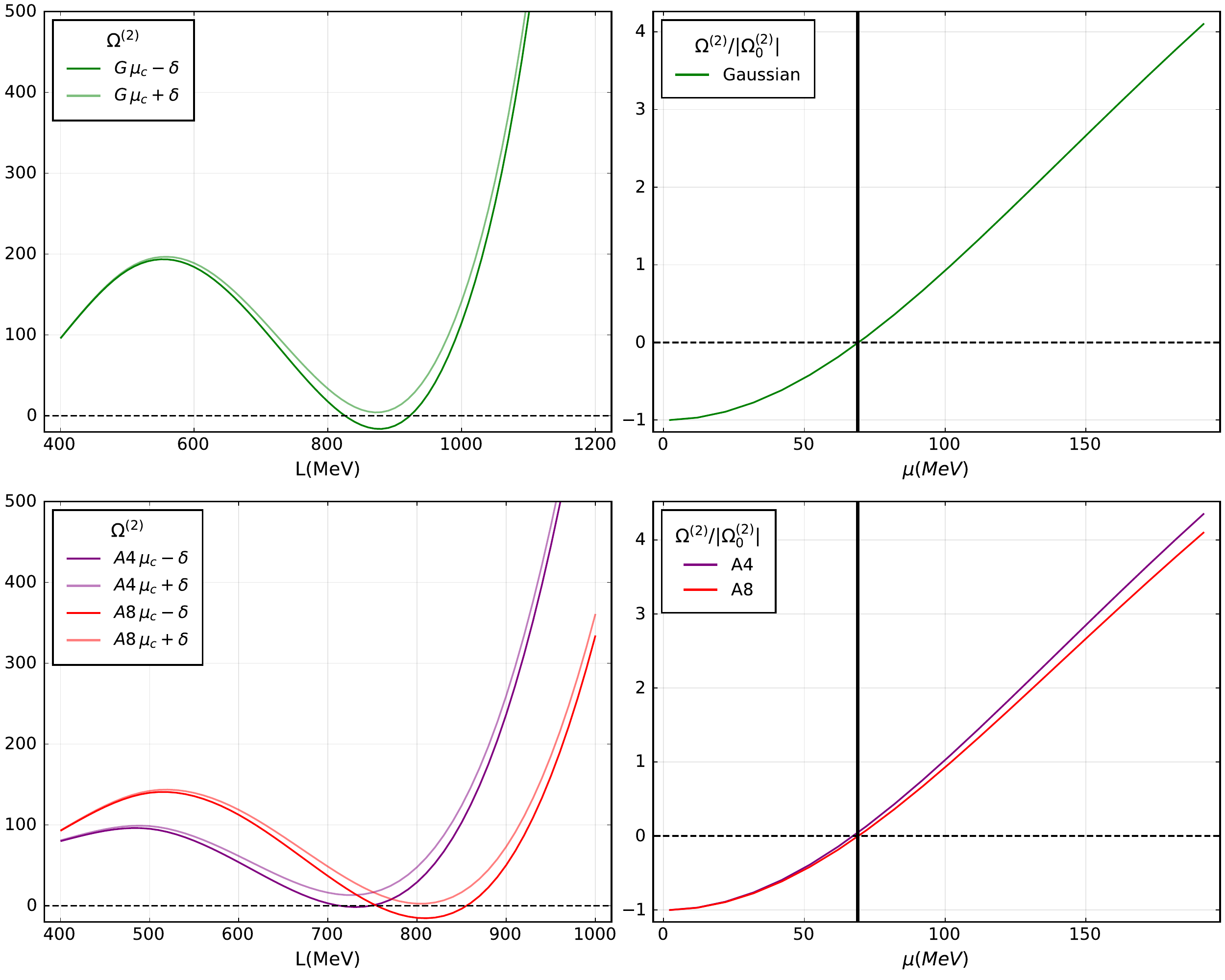}
		\caption{Stability analysis for the homogeneous chiral phase transition test. Left diagrams: thermodynamic potential at a fixed
			value for temperature $T=130\text{MeV}<T_c(\mu=0)$ and chemical potential $\mu = \mu_c(T) \pm \delta = 69\text{MeV} \pm 1\text{MeV}$ as a function of the scale parameter $L$. For the upper
			diagram we used the Gaussian test-function for $\delta m(k)$ and for the lower diagram the algebraic one, with N=4 (A4) and with N=8 (A8), see main text for further explanations.
			Right diagrams: normalised thermodynamic potential as a function of chemical potential on the same fixed temperature slice $T<T_c(\mu=0)$. Note that, although the minimal $L$ value changes with chemical potential, the figures on the r.h.s. were calculated with a fixed L, namely, the minimum L close to the critical chemical potential.}
		\label{fig:chiraltestsa}
	\end{figure}
	
	\subsection{Stability Analysis}
	We now perform the calculation of $\Gamma^{(2)}[\delta S]$ on fixed temperature slices using  
	\begin{equation}
	\delta S(k) = \frac{ \delta m(k)}{\vk^2A(k)^2+(k_4+i\mu)^2C(k)^2} \doteq \frac{\delta m (k)}{d(k)}
	\end{equation}
	with denominator $d(k) = \vk^2A(k)^2+(k_4+i\mu)^2C(k)^2$ for several different test-functions for the mass which will be 
	discussed shortly. The final expression for $\Gamma^{(2)}$ is very 
	simple
	\newcommand{\x}{{\resizebox{4pt}{4pt}{\times}}}
	\begin{equation}\label{eq:ct_gamma2}
	\Gamma^{(2)}[\delta m] = V_4\SumInt_k\left(
	-4  N_c \frac{\delta m(k)^2}{d(k)}
	+12 N_c C_F Z_2^2\SumInt_q 
	\frac{\delta m(k)}{d(k)}
	\frac{\delta m(k-q)}{d(k-q)}D(q)
	\right).
	\end{equation}
	We calculate the thermodynamic potential contribution, Eq.~(\ref{TDstability}) and, 
	since the relevancy is only on the sign (negative meaning unstable, positive meaning stable), we normalise 
	it for better visualisation:
	\begin{equation}
	\text{stability criterium: sign of}\quad \frac{\Omega^{(2)}_\mu}{|\Omega^{(2)}_{\mu=0}|}=
	-\frac{\Gamma^{(2)}_\mu}{|\Gamma^{(2)}_{\mu=0}|}.
	\end{equation}

	What remains to be specified is the mass function $\delta m (k)$. While in a general instability analysis the specific form of the 
	energetically favoured state is not known, here in our specific example the situation is different. From our explicit solutions of the
	quark-DSE in the chirally broken phase we know that the resulting mass function is roughly given by a Gaussian. This is a direct consequence
	of our choice of the effective running coupling, Eq.~(\ref{watson}). For the sake of our proof of concept, however, we will pretend
	to not know the specific shape, but test two different ansaetze for $\delta m (k)$. One is 
	the expected Gaussian, the other a polynomial. Both ansaetze contain a free parameter, $L$, which correspond to an intrinsic scale.
	The functions are given by\footnote{Here we fix the real/imag parts $\lambda = (\lambda_1 + i\lambda_2)$ via Eq.~(\ref{eq:MBFconjecture}) and force Eq.~(\ref{adjoint}) by taking the complex conjugate of $\delta m(k)$ for negative frequencies, i.e. $\delta m(-k_4,k)=\delta m(k_4,k)^\star$.}
	\begin{itemize}
		\item Gaussian $\delta m(k) = \lambda  \text{exp}\left(-\frac{k_4^2+\vk^2}{L^2}\right)$
		
		\item Algebraic decaying $\delta m(k) = \lambda \left(1+\frac{k_4^2+\vk^2}{NL^2}\right)^{-N}$
	\end{itemize}
	This choice allows us to test, if we could infer the actual shape and scale of the energetically favoured instability from the stability 
	analysis alone. 
	
	Indeed, we find instabilities only for a range of values of the scale $L$. A typical situation is displayed in the upper left plot of
	Fig.~\ref{fig:chiraltestsa}, where we plot the instability condition for the Gaussian test-functions a function of the scale $L$ for chemical potentials just before and just after the phase transition, respectively at $\mu=68\,$MeV and $\mu=70\,$MeV. Clearly, there is a region around 
	the maximally unstable value $L=880$MeV which is negative and therefore signals our chiral instability, but only for $\mu<\mu_c$. For smaller values of chemical 
	potential, this negative region becomes larger, but never encompasses all possible scales $L$. For any chemical potential $\mu > \mu_c$
	the stability condition is positive for all scales $L$, signalling that the chirally symmetric phase is stable in this region. Plotting
	the stability condition for $L=880$MeV on the upper right diagram of Fig.~\ref{fig:chiraltestsa} as a function of chemical potential
	we can clearly identify the phase boundary at $\mu=\mu_c$. 
	
	Also for the algebraic ansaetze with power $N=4,...,8$ we are able to identify an ($N$-dependent) optimal scale $L$
	(the minima approaches 880MeV for larger N) from the lower left diagram of Fig.~\ref{fig:chiraltestsa} at the same chemical
	potentials. Observe that the minima become
	lower and lower for $N \rightarrow \infty$. The corresponding stability condition as a function of chemical potential,
	displayed in the lower right diagram of Fig.~\ref{fig:chiraltestsa} shows that negative minima do occur for all powers of $N>3$
	at smaller chemical potential and with increasing power $N$ converge to the one of the Gaussian
	test-function in the diagram above. Thus, if we would not have considered the Gaussian test-function in the first place, this asymptotic 
	behaviour would have given us a clue as to try a Gaussian next. Therefore even in the absence of concrete knowledge of the shape
	of the stable solution of the propagator, we are able to extract corresponding information from a series of instability analyses
	using different ansaetze. In the case at hand, this would lead us to the correct Gaussian shape of $\delta m (k)$ (recall again 
	that the mass function of the chirally broken solution of the DSE \textit{is} Gaussian in this truncation).
	\begin{figure}[t]
		\centering
		\includegraphics[width=0.8\linewidth]{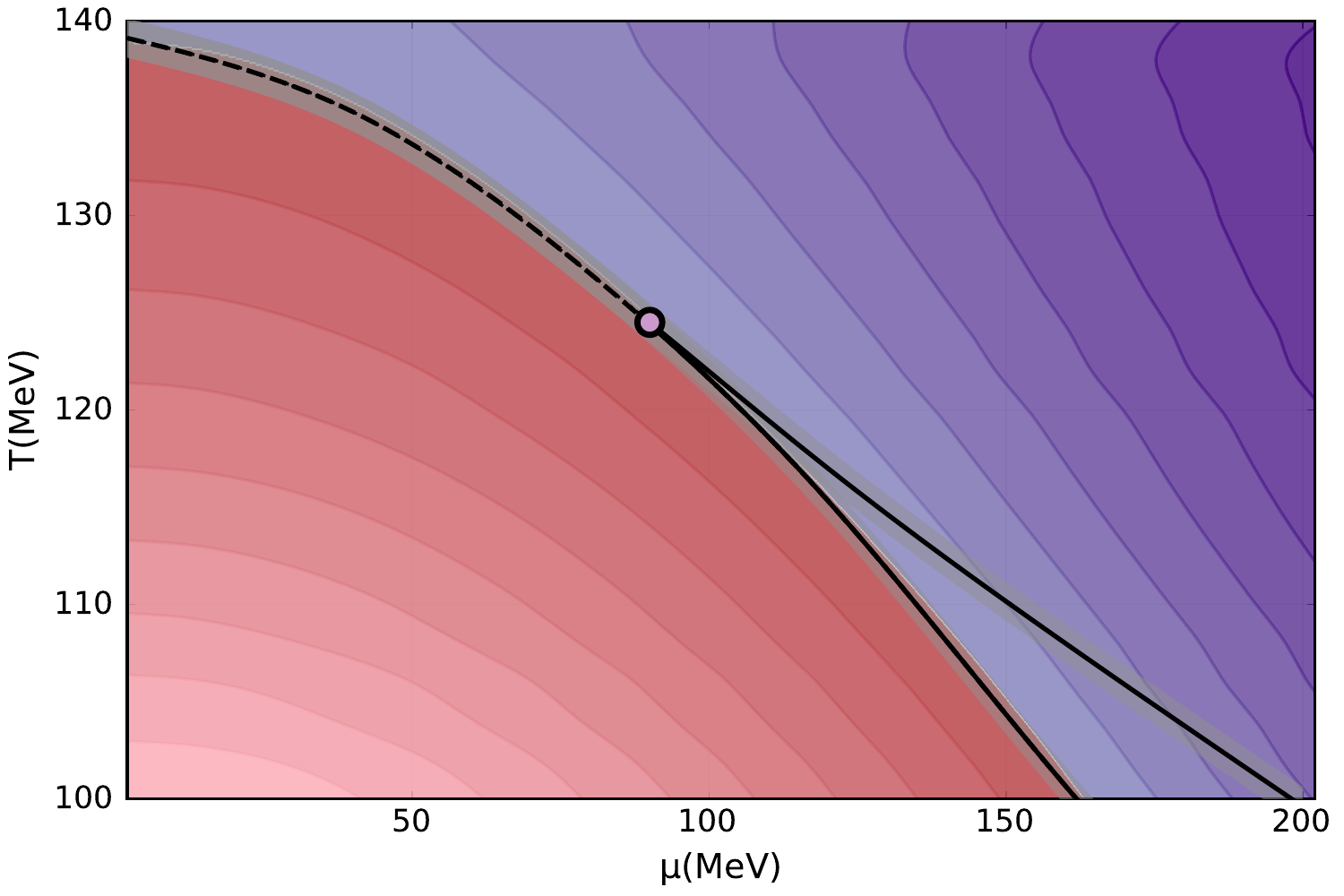}
		\caption{Instability analysis of the chiral Wigner-Weyl solution. Red values indicate negativity (unstable), whereas purple indicates positivity (stable). The colour-gradient scale corresponds to $\Omega^{(2)}$ in code units. Although only the sign of $\Omega^{(2)}$ is relevant for the analysis, naturally small numbers correspond to a shallow stability well.}
		\label{fig:moneyplot}
	\end{figure}
	
	Finally, taking the Gaussian test-function, and performing the stability analysis within the full phase diagram of Fig.~\ref{fig:pdslice}, we obtain the result shown in Fig.~\ref{fig:moneyplot}. The red region depicts where we find $\Omega^{(2)}$ to be negative. Note that the method is not only able to detect the second-order chiral transition line
	but also the left-hand spinodal of the first-order transition. This Wigner-Weyl spinodal line is notoriously difficult to accurately pinpoint numerically, since the calculation of $\Omega^{(2)}(\Gamma^{(2)})$ itself can also be somewhat numerically intricate. The two terms in Eq.~(\ref{eq:ct_gamma2}) are equally large in absolute value, but carry different signs\footnote{Due to this, note the importance of $Z_2$  in Eq.~(\ref{eq:ct_gamma2}). With this model for the gluon $Z_2$ is indeed very close to 1, nevertheless, it drastically change the position of the critical line.}. Therefore, numerically, Eq.~(\ref{eq:ct_gamma2}) corresponds to a subtraction of two big numbers resulting in something several orders of magnitude smaller, which is numerically delicate. Nevertheless, despite these technical difficulties, the method does not seem to ever \textit{overestimate} the unstable region. This is extremely advantageous and increases confidence that whatever instabilities are found, they correspond to true instabilities and not inaccuracies of the numerical methods implemented.

	\section{Conclusions}
	In this work, we have proposed a novel framework for a stability analysis based on the nPI effective action of a given
	quantum field theory. We identified the well-known NJL stability analysis as the 1PI case of this more general framework. 
	Within our approach, we intensively discussed a 2PI-based stability analysis that is suitable for QCD and beyond 
	mean field NJL and QM models.
	
	Although the QCD case is more complex than the one for models, it is possible to arrive at meaningful results.
	We performed a first cross-check of our approach for the case of the homogeneous chiral phase transition, where 
	also explicit numerical results for both, the chirally broken and the chirally symmetric phase are present. We showed 
	that our approach correctly identifies those areas of the phase diagram, where the chiral solution is unstable towards 
	the broken one. We also showed that different test functions lead to a similar result. 
	
	The analysis as it stands is in principle ready to be applied to the search for inhomogeneous phases. One necessary first step for such an analysis, is the determination of the appropriate test-functions. As alluded to above, the form of Eq.~(\ref{deltaSchiral}) was already known \textit{a priori}, which is not the case for inhomogeneous phases. Furthermore, the calculation 
	of the 2PI effective action part $\Gamma^{(2)}$ (or, equivalently, the thermodynamic potential $\Omega^{(2)}$) is far more numerically intensive in this case. In addition, one must perform a search over $\mu$ and $T$ to 
	find the phase transition lines, a task that was greatly alleviated in our cross-check by beforehand knowledge of 
	the transition line. Moreover, optimisation procedures of scales (like $L$ in our case discussed above) in the test 
	function have to be carried out at every analysed point in the ($T$,$\mu$)-plane. This makes the search for inhomogeneous 
	phases a viable but highly cost-intensive procedure that is relegated to future work.
	
	\section*{Acknowledgements}
	We thank Wilhelm Kroschinsky, Ricardo Costa de Almeida and Jesuel Marques for immensely helpful discussions.
	This work has been supported by the Alexander von Humboldt Foundation, the Deutsche Forschungsgemeinschaft 
	(DFG) through the Collaborative Research Center TransRegio CRC-TR 211 
	``Strong-interaction matter under extreme conditions'' and the individual grant FI 970/16-1, the 
	Helmholtz Graduate School for Hadron and Ion Research (HGS-HIRe) for FAIR and the GSI Helmholtzzentrum f\"{u}r
	Schwerionenforschung. 
	
	\appendix
	
	\section{Fourier conventions}
	\label{appendixFourier}
	
	Consider the quark propagator $S(x,y)$ in Euclidean space at finite temperature, 
	which in general is a function of two independent coordinates
	$x,y \in V_4 =  \mathbb{R}^3 \times [0,\frac{1}{T}]$.
	Suppressing Dirac and other discrete indices we define the inverse propagator $S^{-1}$ by
	\begin{equation}
	\label{SinvS}
	\int_z S^{-1}(x,z) S(z,y) = \delta(x,y) ,
	\end{equation}
	where integrals in coordinate space are always understood to be over $V_4$.
	
	The correct fermionic boundary conditions are implemented to these functions by relating them to their Fourier modes
	$S(p,p')$ and $S^{-1}(p,p')$, respectively, as
	\begin{equation}
	S^{(-1)} (x,x') =   \SumInt_{p,p'} e^{-i(p\cdot x- p'\cdot x')} S^{(-1)}(p,p')
	\end{equation}
	with $ \SumInt$ as defined in \Eq{sumint}.
	The corresponding inverse Fourier transforms are given by
	\begin{equation}
	S^{(-1)} (p,p') = \int_{x,x'} e^{i(p\cdot x- p'\cdot x')} S^{(-1)} (x,x') ,
	\end{equation}
	and one can easily check that the following relation holds:
	\begin{equation}
	\SumInt_{k} S^{-1}(p,k) S(k,p') = \int_z e^{i(p-p')\cdot z} 
	= (2\pi)^3 \delta(\vp -\vp\,') \frac{1}{T} \delta_{n,n'},
	\end{equation}
	where $n$ and $n'$ label the Matsubara frequencies corresponding to $p$ and $p'$, respectively.
	
	In {\it homogeneous} matter the propagator only depends on the difference of the coordinates, $S^{(-1)}(x,y) = \bar{S}^{(-1)}(x-y)$.
	We then define
	\begin{equation}
	\bar{S}^{(-1)}(x-y) = \SumInt_p e^{-ip\cdot(x-y)} \bar{S}^{(-1)}(p) ,
	\end{equation}
	which is consistent with \Eq{SinvS} for $\bar S^{-1}(p) \bar S(p) = 1\!\!1$.
	
	In general self-energies $\Sigma = S^{-1} - S_0^{-1}$ are transformed in the same way as the inverse propagators.
	For {\it local} self-energies  $\Sigma(x,y) \equiv \Sigma(x)\delta(x,y)$ (see \Eq{localsigma}) we obtain
	\begin{equation}
	\Sigma(p,p') = \int_x e^{i(p-p')\cdot x} \, \Sigma(x)  \equiv \Sigma(p-p') 
	\end{equation}
	with the inverse transformation
	\begin{equation}
	\Sigma(x) = \SumInt_q e^{-iq\cdot x} \,\Sigma(q) .
	\end{equation}
	Considering space dependent but time-independent local self-energy fluctuations $\delta\Sigma(x) \equiv \delta\Sigma(\vx)$
	we then get 
	\begin{equation}
	\delta\Sigma(q) = \int_{x_4} e^{iq_4 x_4} \int_{\vx} e^{i\vq\cdot\vx}\,\delta\Sigma(\vx)  
	= \frac{1}{T} \delta_{q_4,0} \, \delta\Sigma(\vq)
	\end{equation}
	with 
	\begin{equation}
	\delta\Sigma(\vq) =  \int_{\vx} e^{i\vq\cdot\vx}\,\delta\Sigma(\vx).
	\end{equation}
	
	
	\section{Derivatives}\label{appendixder}
	Take a certain operator defined by a trace of some (continuous and dense) matrices in configuration or momentum space, say $\Gamma = \Tr[MN]$, that is
	$$\Tr[MN] = \int_{x,y} M(x,y) N(y,x).$$
	If we take a derivative, say
	$$
	\frac{\delta \Gamma}{\delta M(x,y)}
	$$
	that is
	$$
	\begin{aligned}
	\frac{\delta \Gamma}{\delta M(x,y)}
	&=
	\int_{x',y'} \frac{\delta M(x',y')}{\delta M(x,y)} N(y',x')
	\\&\hspace{-0.7cm}=
	\int_{x',y'} {\delta(x-x')}{\delta(y-y')} N(y',x') = N(y,x).
	\end{aligned}
	$$
	By that, we note that the derivative of this trace yields not $N(x,y)$ but $N(y,x)$.
	Therefore, when we write
	$$
	\Gamma[S]=-\Tr\log[S]-\Tr[\boldsymbol{1} - S_0^{-1}S] + \Phi_\text{2PI}
	$$
	the stationary point $\delta\Gamma/\delta S(y,x)=0$ is
	\begin{equation}
	\frac{\delta \Gamma}{\delta S(y,x)} =
	- S(x,y)^{-1} - S_0(x,y)^{-1} + \frac{\delta \Phi_\text{2PI}}{\delta S(y,x)} = 0
	\end{equation}
	and thus we identify the self-energy
	$$\frac{\delta \Phi_\text{2PI}}{\delta S(y,x)} = \Sigma(x,y)$$
	and the DSE
	\begin{equation}
	S(x,y)^{-1} = S_0(x,y)^{-1} - \Sigma(x,y).
	\end{equation}
	
	\section{Connected $n$-point Functions}\label{appendixnpoint}
	Consider a theory with a classical action $S[\varphi]$. For simplicity we consider it as a function of a single field $\varphi$, however, the generalisation is trivial. We can write its Euclidean-space generating functional, including $n$ source terms for each $n$-point function, as simply
	\begin{equation}
	Z[J,R^{(2)},R^{(3)},\cdots] = \int \mathcal{D}\varphi \, e^{-(S[\varphi] - J_i\varphi_i - R^{(2)}_{ij}\varphi_i\varphi_j - R^{(3)}_{ijk}\varphi_i\varphi_j\varphi_k - \cdots)}.
	\end{equation}
	It is well known that the derivatives of such a functional with respect to the sources at the physical point are the $n$-point functions.
	\begin{equation}
	\frac{\delta^n Z}{\delta J_{i_1} \cdots \delta J_{i_n}}=(-1)^n
	\frac{1}{Z[0]} \int \mathcal{D}\varphi \, \varphi_{i_1} \cdots \varphi_{i_n} \, e^{-(S[\varphi] - J_i\varphi_i - R^{(2)}_{ij}\varphi_i\varphi_j - R^{(3)}_{ijk}\varphi_i\varphi_j\varphi_k - \cdots)}\Bigg|_{J,R\rightarrow 0}.
	\end{equation}
	However, one can also obtain higher-order $n$-point functions via differentiation with respect to the higher-order source terms. That is, for instance, the four-point function
	\begin{equation}
	\frac{\delta^4 Z}{\delta J_{i_1}\delta J_{i_2}\delta J_{i_3}\delta J_{i_4}}
	=
	\frac{\delta^2 Z}{\delta R^{(2)}_{i_1 i_2}\delta R^{(2)}_{i_3 i_4}}
	=
	-\frac{\delta Z}{\delta R^{(4)}_{i_1 i_2 i_3 i_4}}=\avg{\varphi_{i_1}\varphi_{i_2}\varphi_{i_3}\varphi_{i_4}}.
	\end{equation}
	The connected generating functional is defined as
	\begin{equation}
	W[J,R^{(2)},R^{(3)},\dots] = \log Z[J,R^{(2)},R^{(3)},\dots].
	\end{equation}
	and differentiation of which will give rise to the connected $n$-point functions in the standard way \cite{rivers1988path,peskin2018introduction}, in particular for the 4-point function,
	\begin{equation}
	\frac{\delta^4 W}{\delta J_{i_1}\delta J_{i_2}\delta J_{i_3}\delta J_{i_4}}
	=
	\frac{\delta^2 W}{\delta R^{(2)}_{i_1 i_2}\delta R^{(2)}_{i_3 i_4}}
	=
	-\frac{\delta W}{\delta R^{(4)}_{i_1 i_2 i_3 i_4}}=\avg{\varphi_{i_1}\varphi_{i_2}\varphi_{i_3}\varphi_{i_4}}_\text{connected}.
	\end{equation}
	From such, we may define its Legendre transform, the effective action
	$$\Gamma[\phi,S_2,S_3,\dots].$$
	
	\bibliographystyle{ieeetr}
	\bibliography{stabilitybib.bib}
	
\end{document}